\documentclass[12pt]{iopart}
\usepackage[colorlinks, citecolor = blue, urlcolor = blue]{hyperref}

\usepackage{graphicx}
\usepackage{caption}
\usepackage{subcaption}
\usepackage{bm}

\usepackage{orcidlink}

\begin{document}


\title[An iterative deep learning procedure]{An iterative deep learning procedure for determining electron scattering cross-sections from transport coefficients} 



\author{Dale L Muccignat\orcidlink{0000-0002-6693-6851}$^1$, Gregory G Boyle\orcidlink{0000-0002-8581-4307}$^1$, Nathan A Garland\orcidlink{0000-0003-0343-0199}$^{2,3}$, Peter W Stokes\orcidlink{0000-0002-0956-5927}$^{1,4}$ and Ronald D White\orcidlink{0000-0002-0956-5927}$^1$}
\address{$^1$College of Science \& Engineering, James Cook University, Townsville, QLD 4814, Australia}
\address{$^2$Centre for Quantum Dynamics, Griffith University, Nathan, QLD 4111, Australia}
\address{$^3$School of Environment and Science, Griffith University, Nathan, QLD 4111, Australia}
\address{$^4$Department of Medical Physics, Townsville University Hospital, Townsville, QLD 4814, Australia}
\ead{\mailto{dale.muccignat@my.jcu.edu.au}}


\date{\today}

\begin{abstract}
    We propose improvements to the Artificial Neural Network (ANN) method of determining electron scattering cross-sections from swarm data proposed by coauthors. A limitation inherent to this problem, known as the inverse swarm problem, is the non-unique nature of its solutions, particularly when there exists multiple cross-sections that each describe similar scattering processes. Considering this, prior methods leveraged existing knowledge of a particular cross-section set to reduce the solution space of the problem. 
    To reduce the need for prior knowledge, we propose the following modifications to the ANN method.
    First, we propose a Multi-Branch ANN (MBANN) that assigns an independent branch of hidden layers to each cross-section output. We show that in comparison with an equivalent conventional ANN, the MBANN architecture enables an efficient and physics informed feature map of each cross-section.
    Additionally, we show that the MBANN solution can be improved upon by successive networks that are each trained using perturbations of the
    previous regression. Crucially, the method requires much less input data and fewer restrictive assumptions, and only assumes knowledge of energy loss thresholds and the number of cross-sections present. 
\end{abstract}

\maketitle 


\section{Introduction}\label{sec:Introduction} 

    Electron transport models are crucial to enable the predictive control of low-temperature plasma
    systems~\cite{Adamovich2017}. Underpinning these techniques are the use of accurate and complete electron
    scattering cross-section sets. The derivation of scattering cross-sections is typically
    conducted through experimental and theoretical techniques, coupled with verification through
    swarm scattering experiments to ensure their validity~\cite{Brunger2002}. In regions where these techniques are
    limited, ``educated guesses'' and numerical techniques are often used to
    bridge the gap. This knowledge gap motivates the need for reliable and benchmarked numerical
    techniques to aid in the development of accurate and complete cross-section sets.

    Here, we focus on the determination of cross-section sets from swarm data, otherwise known as the
    inverse swarm problem~\cite{Stokes2020a}. Presently, two primary numerical techniques exist that aim to solve
    the inverse swarm problem: the iterative swarm technique and, more recently, the application of
    Artificial Neural Networks (ANNs). The iterative swarm technique first used approximate
    distributions, such as a Maxwellian or
    Druyvesteyn distribution, of the electron energy distribution function to calculate transport coefficients that are compared to
    experimental transport coefficients and improved iteratively~\cite{Mayer1921,Ramsauer1921,Townsend1922}. The accuracy of the iterative swarm
    technique was then improved with the inclusion of an accurate electron energy distribution functions derived from the solution of
    the Boltzmann equation~\cite{Frost1962,Engelhardt1963,Engelhardt1964,Hake1967}. 

    A substantial limitation to solutions of the inverse swarm problem lies in its ill-posed nature.
    In particular, the existence of multiple cross-sections that describe similar scattering
    processes, such as similar threshold vibrational modes, results in substantial degeneracy of
    transport data for a given species~\cite{Brunger2017,Stokes2020a}. In this limit, the iterative
    swarm technique relies on the intuition and experience of an expert, which, along with the trial
    and error nature of the approach, results in an inefficient procedure that is difficult to
    reproduce. Several methods that attempt to automate this methodology have been
    proposed~\cite{Duncan1972,OMalley1980,Taniguchi1987,Morgan1991a,Morgan1993,Brennan1993} to
    address this issue. Of interest to this study, is use of ANNs trained on existing
    cross-section data to determine scattering cross-sections for electron transport in
    gases.

    In the early 1990s, Morgan~\textit{et.~al.}~\cite{Morgan1991} first demonstrated a solution to
    the inverse swarm problem through an ANN trained on example cross-sections and their
    associated transport coefficients. Recently, Stokes~\textit{et.~al.}~\cite{Stokes2020a}
    revisited this problem utilising advances in network architecture, model size and available
    cross-section data to improve the network's predictive power. Since then, the method has been
    successfully applied to the improvement of tetrahydrofuran, \(\alpha\)-tetrahydrofurfuryl and
    nitric oxide electron scattering cross-sections~\cite{Stokes2020,Stokes2021,Stokes2021a}.
    Jetly~\textit{et.~al.}~\cite{Jetly2021} evaluated the performance of three
    network architectures for the regression of single electron-scattering
    cross-section and found that a DenseNet architecture resulted in the highest regression accuracy.

    While the application of ANNs to the inverse swarm problem shows great promise, the method is
    constrained due to a number of key limitations. This study aims to address the following two
    limitations. First, we demonstrate that a conventional ANN limits the
    ability of the network to model multiple independent cross-section regressions when compared to
    an equivalent network that uses physics informed parallel branches of densely connected layers. Additionally, as the prediction of multiple cross-sections is limited due to the
    degenerate nature of the inverse swarm problem, we
    propose an iterative procedure to enable the network to incrementally explore the
    solution space of the problem.

    In Section~\ref{sec:ML-technique}, we outline each modification to the network
    architecture and methodology before evaluating their performance using methane as a case study. We
    then summarise the results in Section~\ref{sec:conclusion}.

\section{Artificial neural network regression of cross-section sets}\label{sec:ML-technique}

    The application of ANNs towards determining complete cross-section sets
    through the inversion of macroscopic experimental data has been the focus of a recent project at
    James Cook University~\cite[]{Stokes2020,Stokes2020a,Stokes2021,Stokes2021a,Muccignat2022}.
    While the technique has predominately been used to improve existing cross-section
    sets~\cite[]{Stokes2020,Stokes2021,Stokes2021a}, the determination of complete
    cross-section sets for complex targets remains elusive due to the ill-posed nature of the problem. In this section,
    we present two modifications to the methodology that aim to improve the ability of the network
    to determine complete cross-sections from transport data. 

    First, to aid the network in representing the independent nature of each cross-section, we
    propose a Multi-Branch Artificial Neural Network (MBANN) and compare its performance through a regression of the cross-section set for methane recommended by Biagi~\cite{LXCatBiagi7.1}. To reduce the impact of the non-unique nature of determining cross-section sets with multiple similar
    scattering processes, we then propose an iterative procedure that incrementally explores the solution space by using perturbations of the previous regression. To demonstrate the iterative procedure, we compare the initial regression and the best regression found for methane's cross-section set. 
    
    \subsection{Multi-branch neural network regression}\label{sec:parallel}

        Stokes~\textit{et.~al.}~\cite{Stokes2020a} proposed an ANN where each element of the output corresponds to a single cross-section. The simultaneous prediction of each cross-section ensures that the full set of cross-sections are self-consistent, which ensures an accurate replication of the target swarm transport data. In their investigation of various ANN architectures, Jetly~\textit{et.~al.} used separately trained networks to enforce independent feature maps for each cross-section. The authors state that the simultaneous prediction of multiple cross-sections would force the network to share feature maps across different types of cross-sections and thus severely inhibit its predictive capability. 
        
        Here, we propose a Multi-Branch ANN (MBANN) to bridge the gap between the requirement for self-consistency and the desire for independent feature maps for each cross-section. That is, for each cross-section, there exists an independent block of dense layers that each extend from a single block of dense layers. Each parallel branch is then allowed to develop a feature set specific to a single cross-section while still ensuring each regression is conducted in context of the full cross-section set.

        We utilise a MBANN of the form,
        \begin{eqnarray}
            \sigma^{n}\left(\mathbf{x}\right)=
            \left[
                \mathbf{A}_{5}^{n}\circ\mathrm{mish}\circ
                \mathbf{A}_{4}^{n}\circ\mathrm{mish}
                \circ\mathbf{A}_{3}^{n}\circ\mathrm{mish}\circ
                \mathbf{A}_{2}\circ\mathrm{mish}\circ
                \mathbf{A}_{1}
            \right]
            \left(\mathbf{x}\right),
        \end{eqnarray} 
        where $\mathbf{A}_{i}\left(\mathbf{x}\right)\equiv\mathbf{W}_{i}\mathbf{x}+\mathbf{b}_{i}$ are affine mappings defined by dense \textit{weight} matrices $\mathbf{W}_{i}$ and \textit{bias} vectors $\mathbf{b}_{i}$, and $\mathrm{mish}\left(x\right)=x\tanh\left(\ln\left(1+e^{x}\right)\right)$ is a nonlinear activation function~\cite{Misra2019} that is applied element-wise. The final output, $\sigma^{n}$, then represents the \(n^{\textrm{th}}\) cross-section of interest within a set of \(N\) cross-sections. \(\mathbf{A}_{3}^{n}\), \(\mathbf{A}_{4}^{n}\) and \(\mathbf{A}_{5}^{n}\) form an array of \(N\) parallel branches that each utilise the output of \(\mathbf{A}_{2}\) to independently represent the \(n^{\textrm{th}}\) cross-section. \(\mathbf{b}_3^n\) and \(\mathbf{b}_4^n\) contain 32 elements each, \(\mathbf{b}_5^n\) contains 1 for each output \(n\), while those in the initial two layers contain 128. The \textit{weight} matrices are sized accordingly. 
        
\begin{figure}[htpb]
            \begin{center}
                \includegraphics[width=0.45\textwidth]{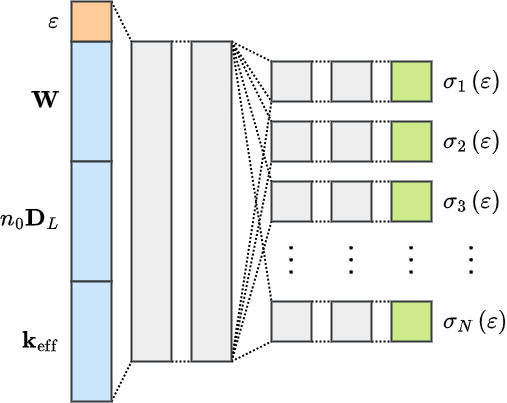}
            \end{center}
            \caption{
                Diagram of the multi-branch artificial neural network used for the regression of cross-sections (green) as a function of energy (orange), given an associated set of transport coefficients (blue). The first two hidden layers contain 128 neurons while each hidden layer within each parallel branch contains 32 (shown not to scale). Each output layer then contains 1 neuron and are concatenated to form an array of N elements to match the number of output cross-sections. }\label{fig:diagram}
        \end{figure}

        From previous investigations and a simple hyper-parameter optimization procedure outlined in \ref{sub:training_procedure}, we found that approximately 32 neurons in each parallel layer is required for a suitable regression of each cross-section with more neurons resulting in modest improvements. Here, we choose this minimum to isolate and demonstrate the differences between a MBANN and an equivalent ANN architecture. Each other parameter, such as the activation function and number of hidden layers, was chosen from a set of reasonable values using a comparison of validation accuracy and prior experience. A schematic representation of the MBANN architecture is shown in Figure~\ref{fig:diagram}.

        For comparison, we use an ANN of the form,
        \begin{equation}
            \sigma\left(\mathbf{x}\right)=
            \left[
                \mathbf{A}_{5}\circ\mathrm{mish}\circ
                \mathbf{A}_{4}\circ\mathrm{mish}\circ
                \mathbf{A}_{3}\circ\mathrm{mish}\circ
                \mathbf{A}_{2}\circ\mathrm{mish}\circ
                \mathbf{A}_{1}
            \right]
            \left(\mathbf{x}\right),\label{eq:neuralnet2}
        \end{equation} 
        where $ \mathbf{b}_3 $ and $ \mathbf{b}_4 $ now contain $ 32\times n $ elements while $ \mathbf{b}_5 $ contains $ N $ elements to match the number of outputs. The ANN thus contains the same number of neurons per layer as the MBANN network outlined above. We note however that the number of trainable parameters is larger than that of the MBANN.
        
        Each cross-section is a function of the incoming projectile electron kinetic energy, $\varepsilon$,
        which, alongside the available swarm data, forms the input to the neural network,
        \begin{equation}
            \mathbf{x}=\left[\begin{array}{c} \varepsilon\\ \mathbf{W}\left( E/n_{0} \right)\\
                    n_{0}\mathbf{D}_\mathrm{L}\left( E/n_{0} \right)\\
                    \mathbf{k}_{\mathrm{\mathrm{eff}}}\left( E/n_{0} \right)\\
            \end{array}\right],\label{eq:input} 
        \end{equation} 
        where \(n_{0}\), \(\mathbf{W}\), \(n_{0}\mathbf{D}_\mathrm{L}\) and \(\mathbf{k}_{\mathrm{eff}}\) are the neutral density, bulk drift velocity, reduced bulk longitudinal diffusion and effective ionisation rate of the electron swarm evaluated at a number of reduced electric fields \(E/n_{0}\). 
        
        To train each neural network, we generate an appropriate set of physically plausible example swarm transport data using augmentations of cross-sections from the LXCat project \cite{Pancheshnyi2012,Pitchford2017,Carbone2021}. The data generation process and subsequent training procedure follows the method outlined by Stokes~\textit{et.~al.}~\cite{Stokes2020a} with some modifications that are described in detail in \ref{sub:data_augmentation}.
        
        A total of \(10^{5}\) training iterations are performed, each consisting of a mini-batch of 32 cross-section sampled at 128 energies from a total of $ 2\times10^{4} $ training examples for each cross-section. For each training set, a multi-term Boltzmann equation solver was used to calculate \(\mathbf{W}\), \(n_{0}\mathbf{D}_\mathrm{L}\) and \(\mathbf{k}_{\mathrm{eff}}\) at 80 log-spaced reduced electric fields between \(10^{-3}\) and \(10^{4}\,\textrm{Td}\) while the cross-section regression was conducted between \(0.01\) and \(200\,\textrm{eV}\). A detailed outline of the training procedure can be found in \ref{sub:training_procedure}.
        
        To demonstrate the improvements offered by the proposed architecture, we present a regression of methane's cross-section set for both the MBANN and an equivalent ANN network. The cross-section set was retrieved from the LXCat database~\cite{Pancheshnyi2012,Pitchford2017,Carbone2021} and originates from Biagi's Magboltz code (version v7.1)~\cite{LXCatBiagi7.1}. In this work, we perform a regression of the elastic momentum transfer, total ionisation, total attachment, and each of the 6 excitation cross-sections.
        In addition, 
        while it has been shown than an ANN can determine some energy loss thresholds~\cite{Stokes2020a}, any target
        cross-section set that exhibits multiple similar threshold processes introduces a high degree of
        degeneracy. In this work, we
        assume knowledge of each energy loss threshold and leave their determination for future
        investigations. 
        
\begin{figure}[htpb]
            \centering
            \begin{subfigure}[htpb]{0.80\textwidth}
                \centering
                    \includegraphics[width=\textwidth]{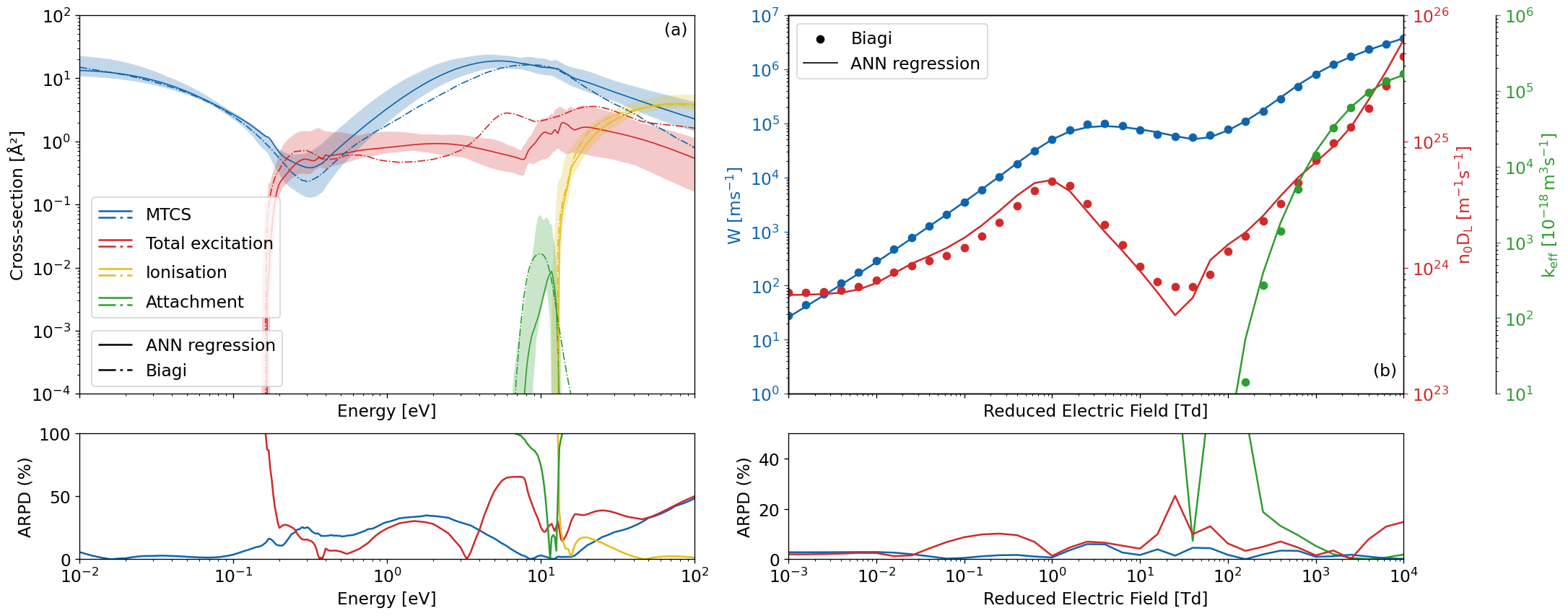}
            \end{subfigure}
            \centering
            \begin{subfigure}[htpb]{0.80\textwidth}
                \centering
                    \includegraphics[width=\textwidth]{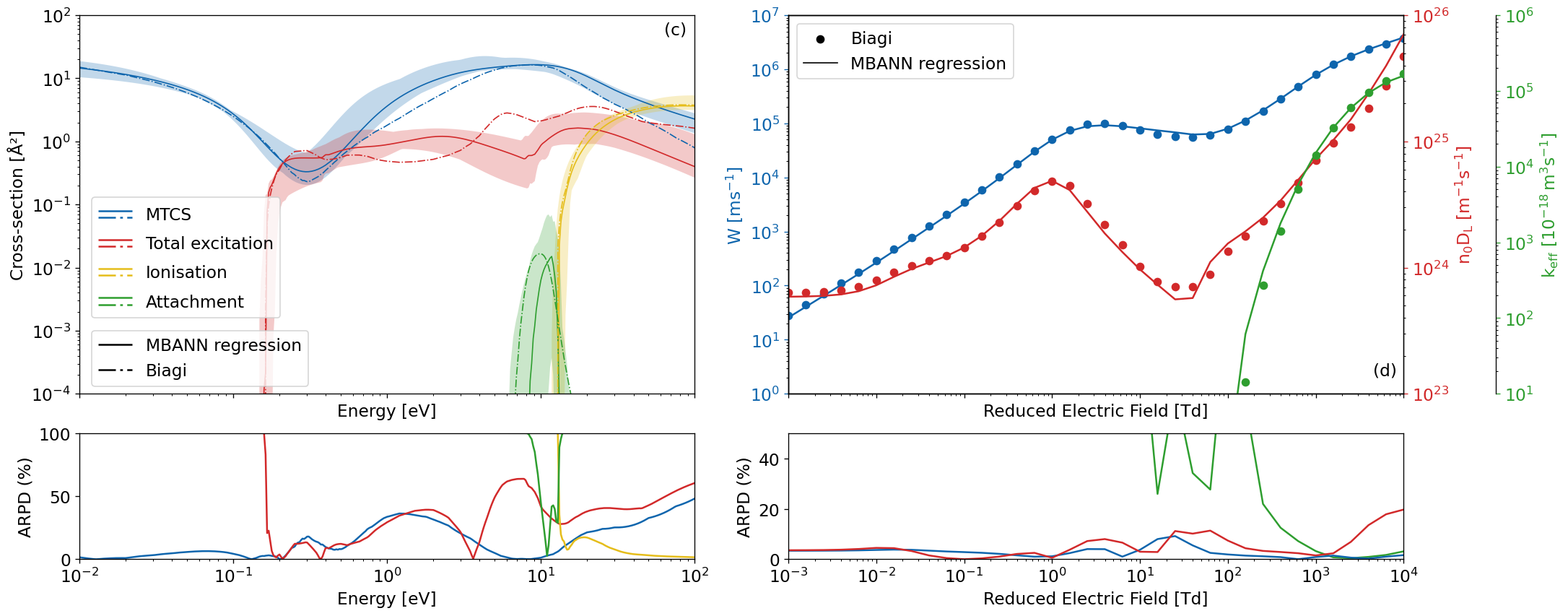}
            \end{subfigure}
            \caption{Comparison of a conventional (a) and an equivalent multi-branch artificial neural network (c) applied to the regression of Biagi's methane cross-section set~\cite{LXCatBiagi7.1}. Shown as shaded regions, are the extend of the best 100 regressions of each network. While only the total excitation cross-section is shown here for simplicity, each of the six excitation processes present in the original set are included in the regression. 
             The bulk drift velocity \(W\), bulk reduced longitudinal diffusion (\(n_0D_\mathrm{L}\)) and the effective ionisation rate (\(k_{\mathrm{eff}}\)) for each network is shown in (b) and (d). All transport coefficients displayed here are calculated with a multi-term Boltzmann equation solver. Below each figure, is the Absolute Relative Percentage Difference (ARPD) across the energy domain between each regression and the original set.}
            \label{fig:ANNvsMBANN}
        \end{figure}

        A comparison between the resulting regression for both the ANN and MBANN architectures is shown in ~Figure~\ref{fig:ANNvsMBANN}. In each, the extent of the 100 best fits sampled during the training process is shown as a shaded region to provide an indication of the network's variability. The ANN regression resulted in a Mean Absolute Relative Percentage Difference (MARPD) of 2.2, 6.3 and \(21\,\%\) for \(W\), \(n_{0}D_\mathrm{L}\) and \(k_{\mathrm{eff}}\), respectively. The MBANN regression resulted in a comparable MARPD of 2.7, 5.2 and \(21\,\%\) for \(W\), \(n_{0}D_\mathrm{L}\) and \(k_{\mathrm{eff}}\), respectively. 
        
        While each network exhibited a similar global accuracy in the replication of transport coefficients, non-physical fluctuations are present in the ANN regression of the elastic and excitation cross-sections between \(0.1-0.4\,\mathrm{eV}\) and \(7-20\,\mathrm{eV}\). It is clear that the large gradients present in these regions, due to the energy loss thresholds of 0.162, 0.363, 7.5, 9.1, 12.36, 15.5 and \(15.5\,\mathrm{eV}\), resulted in the ANN being unable to independently represent each cross-section's feature map when compared to the MBANN fit, despite the same number of neurons available to each architecture. While a larger ANN would be required to mitigate this effect, the MBANN is able to leverage its efficient and intuitive architecture to represent the feature map for each target cross-section.

        As demonstrated by both networks presented here, there remains much room for improvement in the regression of methane's electron-scattering cross-section set. While additional improvements of the network architecture may be available, such as those seen in the work of Jetly~\textit{et.~al.}~\cite{Jetly2021}, the ill-posed nature of the inverse swarm problem places an inherent limit on the accuracy of methods that seek to learn the feature map between transport data and cross-section sets. In what follows, we aim to mitigate this restriction through a new procedure that uses a sequence of MBANNs to incrementally explore the solution space.
        
    \subsection{An iterative approach to neural network regression}\label{sec:iterative}

\begin{figure}[htpb]
            \begin{center}
                \includegraphics[width=0.70\textwidth]{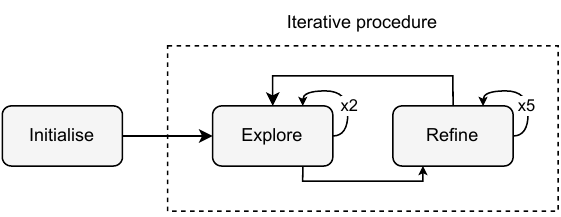}
            \end{center}
            \centering
            \includegraphics[width=0.75\textwidth]{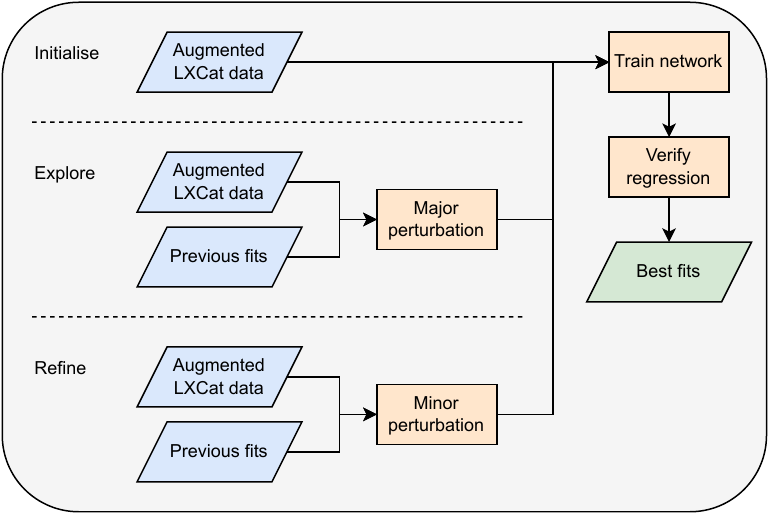}
            \caption{Illustrative diagram of iterative neural network procedure (top) along with a flow chart outlining each step (bottom). The procedure consists of 3 phases: initialise, explore and refine. The initialise step follows a similar methodology outlined by Stokes~\textit{et.~al.}~\cite{Stokes2020a}. First, training data is generated through augmentations of existing LXCat cross-sections~\cite{Pancheshnyi2012,Pitchford2017,Carbone2021}. During training, the network's output is periodically sampled and their associated transport coefficients are verified against the target transport coefficients to determine the best 100 fits. In the explore and refine steps, augmented LXCat data are used to generate major and minor perturbations, respectively, of the previous best 100 fits before utilising the same training and verification procedure as the initialisation step.}\label{fig:Iterative}
        \end{figure}

        The regression of numerous similar cross-sections poses a substantial challenge for solutions to the inverse swarm problem. As shown previously by Stokes~\textit{et.~al.}~\cite{Stokes2020a,Stokes2021,Stokes2021a}, the predictive power of the neural network can be improved by restricting the training data to perturbations around a reference cross-section set. We extend this work and propose an iterative procedure in which a sequence of networks are trained using a weighted mixing of the previous solution with example cross-section data. As illustrated in Figure~\ref{fig:Iterative}, the procedure consists of three phases; initialise, explore and refine. 

        In the intialise phase we follow the procedure outlined in \ref{sub:training_procedure}. In this phase, no prior information of the target cross-section set is given to the network other than energy loss thresholds and the number of processes present. The resulting best 100 regressions then form an array of current fits $ \boldsymbol{\sigma}_c $. During the following two phases, we seek to improve the regression by generating stochastic perturbations around each current fit through a weighted mixing with example cross-section data. 
        
        To train each subsequent network, we use augmented LXCat cross-sections to generate perturbations around each current fit. First, \(\sigma_s\), is generated with the same method used in the initialisation phase. We then use $ \sigma_s $ to generate a perturbation around the \(i^{th}\) current fit \(\sigma_{c,i}\) using a weighted sum in log space, 
        \begin{equation}
            \sigma_{s}\left(\varepsilon\right)=
            \sigma_{s}^{1-r}\left(\varepsilon-\varepsilon_{s}+\varepsilon_{s}^{1-r}\varepsilon_{c,i}^{r}\right)
            \sigma_{c,i}^{r}\left(\varepsilon-\varepsilon_{c,i}+\varepsilon_{s}^{1-r}\varepsilon_{c,i}^{r}\right),
        \end{equation}
        where \(i\) is a uniformly distributed random number and $r$ is a pseudo-random number sampled from a scaled Laplace distribution. The parameter \(r\) then defines how similar each training sample \(\sigma_s\) is to \(\sigma_{c,i}\), where values close to 1 results in minor perturbation around \(\sigma_{c,i}\) while values close to 0 results in major perturbations. Values of \(r\) greater than 1 can be used to produce accentuated perturbations around \(\sigma_{c,i}\) to extend the solution space beyond the available data. The extent of these perturbations then define the network's ability to either explore the solution space or refine the existing solution. If the training data is restricted to minor perturbations, the solution may become trapped in a local minimum. Conversely, major perturbations may result in the network being unable to determine a sufficiently accurate cross-section set. The explore and refine phases of the procedure aim to strike a balance between these two regimes. In the explore phase, major perturbations of \(\boldsymbol{\sigma}_c\) are made to assist the network in traversing the solution space beyond the current fits while in the refine phase, minor perturbations are used to further refine \(\boldsymbol{\sigma}_c\). 
        

        Through a trial and error process, we found the following parameters to be suitable for each training phase. In the explore phase, we conduct two iterations while in the refine phase we conduct five to help ensure sufficient refinement of a particular solution is conducted after each exploratory phase.
        For high energy ($>~10\,\mathrm{eV}$) processes, such as electronic excitation and ionisation, we sample $ r $ from the domain $ \left[0.5,0.8\right] $ for each iteration in the explore phase while during the refine phase we set $ r= 0.8 $. For low energy processes, such as vibration and elastic, $ r $ is sampled from $ \left[0.5,1.5\right]$ in the explore phase and $ \left[0.8,1.2\right] $ in the refine phase. In the case of low energy processes, $ r $ values greater than 1 are used to generate training examples that accentuate low energy cross-section features that are present in the sample data. 
        
        Direct parallels can be drawn between the iterative MBANN technique and the well known iterative
        swarm technique. In each, an informed supervisor guides the procedure towards both a
        physical and accurate solution to the inverse problem. In the iterative swarm technique, this is
        generally the role of an expert in the field who may make adjustments to the solution or procedure where necessary. In the iterative MBANN technique this role is, in the ideal case, automated by the neural network. Depending on the application, the guidance of an expert may still be required to choose suitable parameters and monitor its performance.

\begin{figure}[htpb]
            \centering
            \begin{subfigure}[htpb]{0.90\textwidth}
                \centering
                    \includegraphics[width=\textwidth]{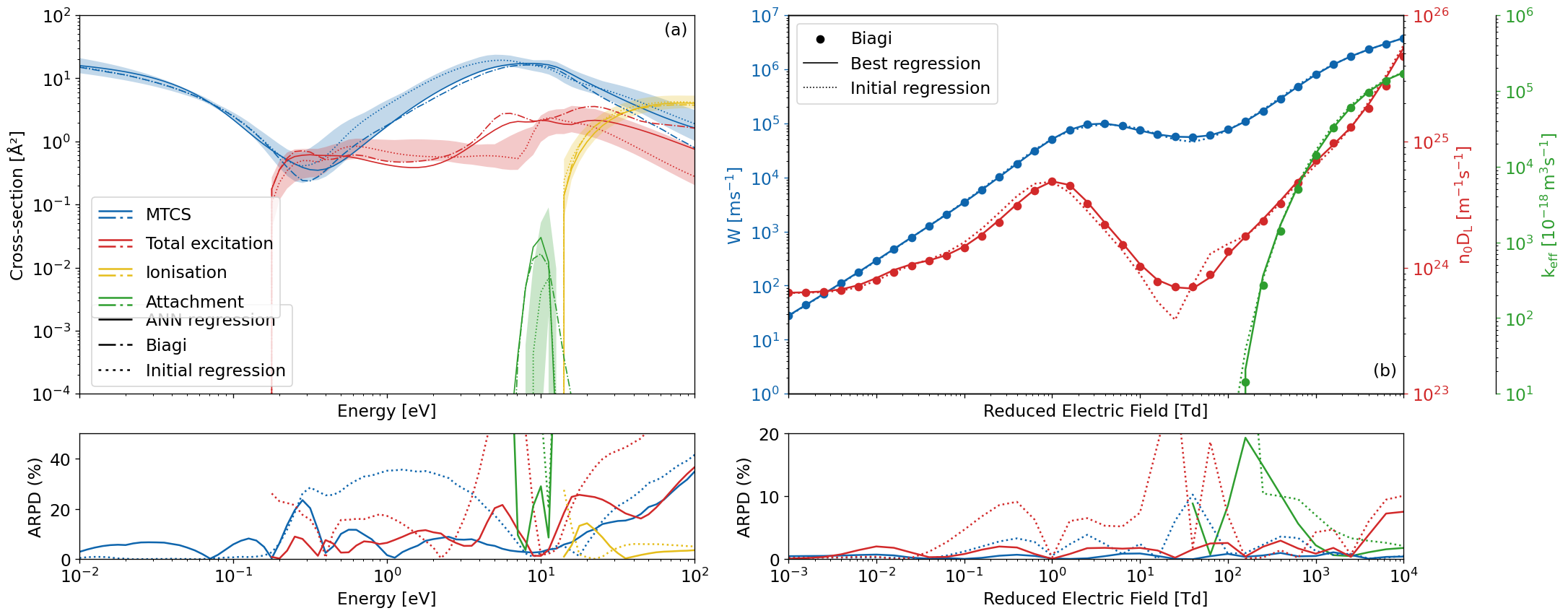}
            \end{subfigure}
            \hfill
            \caption{MBANN regression of Biagi's methane cross-section set~\cite{LXCatBiagi7.1}  using an iterative procedure. (a) compares the initial regression and the best regression found to the original cross-section set.  Shown as shaded regions, are the extent of the best 100 regressions for the initial regression. While only the total excitation cross-section is shown here for simplicity, each of the six excitation processes present in the original set are included in the regression. 
             The bulk drift velocity \(W\), bulk reduced longitudinal diffusion (\(n_0D_\mathrm{L}\)) and the effective ionisation rate (\(k_{\mathrm{eff}}\)) for each regression is shown in (b). All transport coefficients displayed here are calculated with a multi-term Boltzmann equation solver. Below each figure, is the Absolute Relative Percentage Difference (ARPD) across the energy domain between each regression and the original set.} 
            \label{fig:ML-CH4}
        \end{figure}

        We demonstrate the proposed iterative procedure through a regression of methane's cross-section set presented in Section~\ref{sec:parallel}. While 32 neurons was chosen for the hidden layers in each parallel branch as the limiting case in the previous section, we increase this to 64 in what follows due to modest improvements in the validation accuracy. In Figure~\ref{fig:ML-CH4}, we compare both the
        initial and the best regression found of methane's cross-section set during the procedure, along with their associated transport coefficients.

        The associated transport coefficients of the initial fit of methane's cross-section set results in substantial discrepancies to the original set. 
        The initial MBANN regression resulted in a MARPD of 1.8, 5.1 and \(27\,\%\) for \(W\), \(n_{0}D_\mathrm{L}\) and \(k_{\mathrm{eff}}\), respectively. After 40 iterations, the procedure was then able to substantially improve upon the initial regression with the best iteration resulting in a MARPD of 0.48, 1.69, and \(5.84\,\%\) for \(W\), \(n_{0}D_\mathrm{L}\) and \(k_{\mathrm{eff}}\), respectively.

        Crucially, we also find substantial improvements in the agreement between the total cross-sections for each collision type in the set and the target cross-section set. Provided as shaded regions in Figure~\ref{fig:ML-CH4}, is the extent of the best 100 regressions found in the initial fit. Both the total excitation and the attachment cross-section of the best regression exist, in part, outside of the initial extent of \(\boldsymbol{\sigma}_c\). The network was therefore able to effectively explore where necessary to improve the resulting fit of the target transport coefficients.
        
        In its current form, the procedure assumes the prior knowledge of 
        threshold energies. This assumption is particularly important when groupings of similar
        thresholds are present. If instead, effective excitation cross-sections are utilised in the
        network to represent groupings of similar threshold energies, this assumption could be avoided
        at the expense of physical threshold energies in the resulting cross-section set.  
    
        In this investigation, we utilise only calculated transport coefficients over a large range of
        reduced electric fields. In reality, such a range is not often available. While tailoring the
        energy domain of the cross-section regression to the transport data available will
        alleviate the limitation in part, this has limited returns. We thus encourage the measurement of swarm
        coefficients over a broad range of electric field domains where possible. Finally, while we have made a
        concerted effort to develop a robust iterative procedure, depending on the particular problem
        and the extent of available transport data, the network may still produce non-physical
        cross-sections or become trapped in a local minimum. The parameters utilised here should thus
        only serve as a guide for future applications to be modified as needed.

\section{Conclusion} 
\label{sec:conclusion}
    
    In this work, we demonstrate a new iterative procedure that uses a Multi-Branch Artificial Neural Network (MBANN) to solve the inverse swarm problem. Building upon the foundations outlined by
    Stokes~\textit{et.~al.}~\cite{Stokes2020a,Stokes2020,Stokes2021a,Stokes2021} and
    Jetly~\cite{Jetly2021}, we address two key limitations of an ANN solution to the
    inverse swarm problem. 
    
    We first evaluate the use of a MBANN that includes
    an independent branch of dense layers for each output that each stem from common feature
    map of the input. We then compare the MBANN to an equivalent conventional ANN using Biagi's methane
    cross-section set~\cite{LXCatBiagi7.1} and demonstrate that the use of parallel layers can improve the resulting regression as the network is able to efficiently and independently represent multiple distinct cross-sections.

    In addition, taking inspiration from the iterative swarm technique, we propose an iterative MBANN procedure to that incrementally explores the solution space to reduce the ill-posed nature of the problem. After an initial regression is found, we use a sequence of MBANNs that are each trained using perturbations around the previous regression. The iterative MBANN
    procedure then converges towards a particular solution of the inverse swarm problem. 

    To demonstrate the iterative MBANN procedure, we evaluate its performance using Biagi's methane
    cross-section set~\cite{LXCatBiagi7.1}. In the 40 iterations that were conducted, the MARPD of the initial regression's resulting transport coefficients was substantially decreased from 1.8, 5.1 and \(27\,\%\), to 0.48, 1.69,  and \(5.84\,\%\) for \(W\), \(n_{0}D_\mathrm{L}\) and \(k_{\mathrm{eff}}\), respectively. Additionally, the total cross-section for each collision type within the best set found exhibited good agreement with the original set, in contrast to the initial regression. 

    Overall, we have demonstrated an improved artificial neural
    network solution to the inverse swarm problem that utilises both an iterative procedure 
    and parallel branches of densely connected layers that represent each cross-section. In conjunction, these additions improve the ability of the network to generate both
    self-consistent and physical cross-sections, particularly when large degeneracies may exist for a
    particular species. 
    
    In future work, we aim to apply this procedure to derive complete
    cross-section sets for complex targets while also investigating the use of convolutional
    architectures. 

\ack
The authors gratefully acknowledge the financial support of the Australian Research Council (ARC) through the Discovery Projects Scheme (DP220101480 and DP190100696).
\appendix

    \section{LXCat data augmentation} 
    \label{sub:data_augmentation}

        To provide the network with sufficient training samples, we implement the data augmentation
        method first presented by Stokes~\textit{et.~al.}~\cite{Stokes2020a}. First, we collate an appropriate
        set of physically plausible example swarm transport data using cross-sections from the LXCat
        project \cite{Pancheshnyi2012,Pitchford2017,Carbone2021}. Following the method outlined in the
        work of Jetly~\textit{et~al.}~\cite[]{Jetly2021}, from a total of 397 elastic and effective
        cross-sections available, 3 groups of similar features were created by visual inspection. The
        total number of cross-sections contained in each group were 60, 130, and 197. In addition, 6763,
        451 and 332 cross-sections in total were present for excitation, ionisation and attachment
        processes respectively.

        In each iteration, we then generate a database of
        example cross-sections through a randomised mixing of cross-section pairs using a 
        weighted sum in log space,
        \cite{Stokes2020a}:
        \begin{equation}
            \sigma_{s}\left(\varepsilon\right)=\sigma_{1}^{1-r^{\prime}}\left(\varepsilon-\varepsilon_{1}+\varepsilon_{1}^{1-r^{\prime}}\varepsilon_{2}^{r^{\prime}}\right)\sigma_{2}^{r^{\prime}}\left(\varepsilon-\varepsilon_{2}+\varepsilon_{1}^{1-r^{\prime}}\varepsilon_{2}^{r^{\prime}}\right),\label{eq:mixture-mag}
        \end{equation}
        where $\sigma_{1}$ and $\sigma_{2}$ are a random
        pair of LXCat electron scattering cross-sections, sampled from the available targets, for a
        given process type (e.g., excitation, ionisation, etc.), while $\varepsilon_{1}$ and $\varepsilon_{2}$ are their respective threshold
        energies.
        The parameter $r^\prime$ is a pseudo-random mixing ratio. To help ensure a physical representation of cross-sections within the training set, a scaled Laplace distribution truncated to
        the domain $ r\in\left[0,1.5\right] $ was chosen to bias $r^\prime$ towards small perturbations around each example cross-section. Any decaying distribution may be sufficient for this purpose however. Ratios greater than 1 are used here
        to accentuate cross-section features found within the
        sample set to reduce the extend of outliers within the set. Note that for the elastic cross-section, we sample two
        cross-sections from the three groups so that each group is equally
        represented in the training set. 

        In addition, due to the limited nature of the available data, the solution may exist at the
        extremes of the available training data which can introduce unwanted bias in the data
        augmentation process. To alleviate this, Equation~\eref{eq:mixture-mag} is modified such
        that the energy domain and magnitude of each cross-section are multiplied by the scaling factors \(10^a\) and \(10^b\)
        respectively. Each factor is a pseudo-random number uniformly distributed within a defined
        range. Here, we set $ a\in\left[-0.5,0.5\right] , b\in\left[-0.5,0.5\right] $ for elastic cross-sections, $ a=0
        , b\in\left[0,2\right] $ for excitation cross-sections, $ a=0 ,
        b\in\left[0,1\right] $ for ionisation cross-sections and $ a\in\left[-1,1\right] ,
        b\in\left[-1,1\right] $ for attachment cross-sections. Each was chosen
        to reasonably extend the extent of the available training data.
        Finally, we log transform then scale each cross-section between \(-1\) and \(1\). If a
        cross-section magnitude is below \(\delta=10^{-6}\), it is replaced by \(10^{-7}\) before applying
        the transform. 

        For each generated cross-section set, the resulting transport coefficients are calculated
        through a multi-term Boltzmann equation solver according to the target experimental transport
        coefficients. If a particular set results in non-physical transport coefficients, the set is
        removed from the training set and a new set is found until the condition is satisfied. For this
        investigation, physical transport coefficients are defined as \(\mathbf{W}>0\) and
        \(n_{0}\mathbf{D}_\mathrm{L}>0\), where the electric field is directed along the negative z-axis. In
        addition, we apply a logarithmic transformation to ensure that all inputs and outputs of the
        network are dimensionless and lie within $\left[-1,1\right]$, with special consideration 
        given to \(\mathbf{k}_{\mathrm{eff}}\) due to the presence of negative values. The vectors
        \(\mathbf{k}_{\mathrm{eff}}^{+}\) and \(\mathbf{k}_{\mathrm{eff}}^{-}\) are created to represent the positive and
        negative portions of the original input respectively. For \(\mathbf{k}_{\mathrm{eff}}^{+}\), each
        negative value is set to a sufficiently small positive value while for \(\mathbf{k}_{\mathrm{eff}}^{-}\),
        each positive value is set to a sufficiently small negative value before taking the absolute
        value of each. 

    \section{Training procedure} 
    \label{sub:training_procedure}

        Training of the network is conducted using a mini-batch of 32 cross-section sets evaluated at
        128 energies for a particular iteration. Each \textit{weight} and \textit{bias} is then updated
        using the `NAdam' optimiser~\cite{Dozat2016} with a learning rate of \(0.001\) and
        \(0.9\) and \(0.999\) as the exponential decay for the first and the second momentum estimate
        respectively. For each
        batch, random noise is applied to each transport coefficient. Noise is sampled from a
        lognormal distribution with a standard deviation chosen to replicate experimental uncertainties typically observed for each transport coefficient.
        
        The prediction of multiple cross-sections of the same collision type has previously been shown
        to be a highly degenerate problem, particularly in the case of excitation
        collisions~\cite{Stokes2020a}. To emphasise the importance of the total cross-section, we use a loss function that includes a penalty
        for the total cross-section for each process type, in addition to each individual cross-section. The
        loss function is defined as follows,
        \begin{equation}
            L=\frac{1}{I}\sum_i^I\frac{1}{M_{i}+1}\left[
                \sum_{\sigma\in\boldsymbol{\sigma}_{i}}\left|\sigma-\hat\sigma\right| 
                + \left|\sum_{\sigma\in\boldsymbol{\sigma}_{i}} \sigma -
            \sum_{\sigma\in\boldsymbol{\sigma}_{i}}\hat\sigma\right|\right],
        \end{equation}
        where \(\boldsymbol{\sigma}_i\) is the set of $ M_i $ cross-sections for the collision type
        \(i\) while \(I\) is the number of collision types present. Note that the calculation of the
        total cross-sections is conducted in linear space before scaling is re-applied.
        
        To validate the network, we set aside \(5\,\textrm{\%}\) of the available cross-sections for each
        collision type to form the basis of our validation set. In the case of the elastic
        cross-sections, \(5\,\textrm{\%}\) of each of the feature groups is selected. The chosen
        validation cross-sections then undergo the same data augmentation as the training data to
        produce 100 samples. Every 100 iterations we test the network on the
        validation set and calculate the MARPD over the
        energy domain,
        \begin{equation}\label{eq:marpd}
             \textrm{MARPD} = \frac{1}{I}\sum_{i}^I \left| 100\times
             \frac{\sigma_{t,i}-\hat{\sigma}_{t,i}}{\left| \sigma_{t,i}
             \right| + \left| \hat{\sigma}_{t,i} \right| } \right|,
        \end{equation} 
        where \(\sigma_{t,i}\) is the total cross-section for the collision type \(i\). During training, the validation loss is used to both compare network parameters and aid in preventing overfitting.

        In addition, every 10 training iterations we store the neural network's prediction of the
        target transport data and compare the resulting transport coefficients. Due to the presence
        of multiple distinct fit parameters, we repeatedly select one random parameter and then remove the worst
        predicted cross-section set until only one remains. This process is then replicated 1000 times before the cross-section set which
        appears most frequently is chosen as the best overall fit. As desired, the best overall fit can be removed, and the processes repeated to find the next best fit.
        
        To select each hyperparameter, a simple tuning process was conducted that compared
        the validation loss between reasonable values for batch size (32, 64, 128), number of hidden
        layers (2, 3, 4, 5), hidden layer size (32, 64, 128, 256) and the optimiser used (NAdam, Adam). For simplicity, this processes was conduced with a conventional ANN in which each layer contained an equal number of neurons and the output consisted of an elastic and ionisation cross-section along with three electronic excitation cross-sections. The tuning processes indicated that 4-5 hidden layers with 128-256 neurons resulted in the smallest validation error.
        Due to computational limitations, a more extensive tuning that includes branching hidden layers and a greater variation of parameters was not conducted.

\bibliographystyle{iopart-num}
\bibliography{Zotero.bib}
\end{document}